# Numerical Contribution of Thermoelectric Polyaniline Sheet

J. Dgheim[†*], C. Jreich[†], M. Ghazeleh[†], P. Ziadeh[†], M. Abdallah[†]

† Laboratory of Applied Physics (LPA), Group of Mechanical, Thermal & Renewable Energies (GMTER), Lebanese University, Faculty of Sciences II.

* E-mail: jdgheim@ul.edu.lb

**Abstract**: This work concerns the study of thermo-physical properties and transport phenomena in thermoelectric liquid polyaniline sheet. The electro-thermal heat transfer equations coupled to Naviers-Stokes equation, continuity equation, initial and boundary conditions are solved using the finite difference and the finite element schemes. The results of both numerical techniques show good qualitative and quantitative agreements. The voltage difference, the temperature variation, the Seebeck coefficient, the figure-of-merit and the maximum efficiency of polyaniline sheet are determined numerically. Our results show that the voltage difference is equal to 0.09 mV for non doped polyaniline and 0.20 mV for doped polyaniline DPAN/HCl(I) for a sheet dimension of 4×4×12 mm$^3$.

## 1. Introduction

Since the industrial revolution, global electricity consumption has been growing exponentially. The global demand for fossil fuels and the increase in energy needs are declining these energies. In addition, these fossil fuels are responsible of climate change [1,2]. Due to the impact of energy production on the environment, the need to reduce greenhouse gas emissions and the depletion of fossil fuels, the energy policies of the government have been directed towards renewable energies. Thermoelectricity is one of the solutions for converting large amounts of waste heat into electrical energy. Nowadays, the thermoelectric materials can play a very important role in converting directly heat into electricity [3,4]. The active part of these devices consists of conductive materials and has the advantage of having a quiet and reliable system. The physical principle of the thermoelectric materials is associated to three fundamental effects [5,6]:

- First, the Seebeck effect is referred to be the thermo-power effect. The electrical potential is produced within a single conductor when a temperature gradient is subjected.
- Second, the Peltier effect, is produced when a difference of temperature is created at the junctions of two dissimilar conductors while an electric current is applied.
- Third, the Thomson effect, in which the heat contained in a single conductor is changed into a gradient of temperature while an electric current passes through it.

The energy conversion efficiency of thermoelectric materials is quantified by the dimensionless figure-of-merit *ZT*. Best of thermoelectric material should have:

- a high electrical conductivity to minimize Joule heating,
- a low thermal conductivity to prevent thermal shorting,
- a large Seebeck coefficient for the maximum conversion of heat to electrical power.

Many materials with low thermal conductivity have been investigated [7-9]. Reducing the thermal conductivity improves the thermoelectric figure of merit coefficient. Kumar et *al*. [10] developed a numerical model to simulate coupled thermal and electrical energy transfer processes in a thermoelectric generator designed for automotive waste heat recovery systems. Shi et *al*. [11] has developed a two-dimensional model to investigate the performance of thermoelectric module. The voltage and temperature distributions of the model under two types of boundary conditions (constant cold-side temperature and fixed convection heat transfer coefficient) are studied. Semiconductors materials have good thermoelectric performance, and they can be classified according to their temperature efficiencies. For the applications around the ambient temperature, several alloys are used (Sb$_2$Te$_3$-p type or Bi$_2$Se$_3$-n type). The tellurium and bismuth alloy (Bi$_2$Te$_3$) can be considered as competitive. For temperatures above 450 K, The lead tellurium (PbTe) compound presents acceptable efficiency. At high temperatures (>1000 K), n and p types of Silicon-Germanium alloys (SiGe) are very efficient [12].

On the other hand, complex thermoelectric liquids are also very promising materials. Compared to semiconductor thermoelectric, thermoelectric liquids have many advantages (less toxic and inexpensive). In addition, these liquids have high Seebeck coefficients (greater than 1 mV/K) [13-14]. Several thermoelectric liquids have been studied [15], including ionic liquids formed from cations (dialkylimidazolium, tetraalkylammonium, ...) and anions (nitrate, ...), aqueous electrolytes (Ferri/potassium ferrocyanide solution, ...), and non-aqueous electrolytes (tetrabutylammonium nitrate, ...). The Seebeck effect in liquid thermoelectric materials describes the difference in particle concentration in a complex fluid. It is related to the difference of ionic transport of positive and negative ions and it corresponds to the appearance of a voltage between the two faces of a material when a temperature difference is applied between them. The applied temperature gradient generates a migration of positive and negative ions from one face to another [16].

Our work is focused on the study of thermo-physical properties and transport phenomena in thermoelectric liquids of doped and non-doped polyaniline PANI. This last is a very promising liquid with interesting properties for thermoelectric applications.

## 2. Mathematical Model

The heat transfer in thermoelectric materials modifies the internal energy of the materials. The rate of the heat transport depends on the temperature, and the properties of the medium, through which the heat is developed. Our physical model consists of a two-dimensional square cell filled with a thermoelectric liquid (Figure 1). Two sources of heat are placed on both sides of the cell. They are used to adjust the temperatures of the two opposite sides. The hot temperature $T_h$ is fixed on the part tangent to the $y$ axis, while the cold temperature $T_c$ is fixed on the part parallel to the $y$ axis, assuming that the sides parallel to the $x$ axis of the cell are adiabatic. The $x$ and $y$ axes are tangent to the cell and $O\ (0,0)$ is considered the origin of the system. The module receives a certain amount of heat from the hot source.

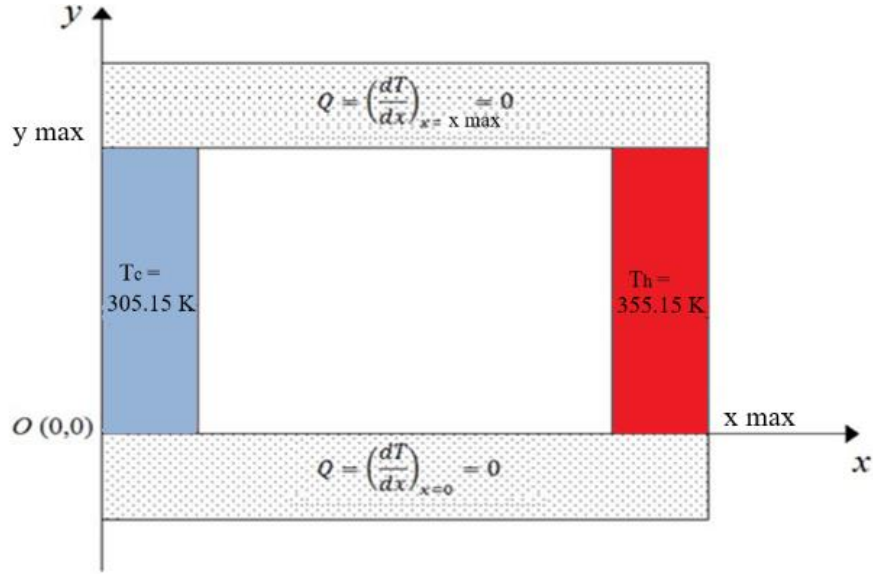

**Figure 1:** Physical model.

This supplied thermal power is transformed into work by the thermoelectric module in the form of an electric current of output *I*. The thermal transfer enables to increase the entropy of the system, thus increasing the Seebeck coefficient. The mathematical model assumes laminar and steady flow and neglects chemical reaction, radiation, and Soret and Dufour effects. The ambient pressure is the atmospheric pressure. Using the simplifying conditions, the thermoelectric model is summarized by the heat transfer equation coupled with Seebeck, Thomson and Peltier effects, as well as the Naviers-Stokes and continuity equations. The thermo-electric transfer equation is written as the following:

The simplified heat equation:

$$\rho C p \frac{\partial T}{\partial t} = \frac{\partial}{\partial x}\left(K \frac{\partial T}{\partial x}\right) + \frac{\partial}{\partial y}\left(K \frac{\partial T}{\partial y}\right) + \frac{J^2}{\sigma_T} - \tau \vec{J} \nabla T \qquad (1)$$

Where $\rho$ is the density, $C_p$ is the specific heat at constant pressure, $T$ is the temperature, $t$ is the time, $K$ is the thermal conductivity, and $\sigma_T$ is the electrical conductivity at constant temperature. The term $\tau$ is the Thomson coefficient and $J$ is the electrical density both considered as variable in space and time.

The Peltier-Thomson term is the following: $T\vec{J}\nabla \alpha$ (2)

Equation (2) contains both Thomson and Peltier contributions. By using the equivalence second relationship of Kelvin $\pi = \alpha T$, one can obtain:

$$T\vec{J}\nabla\alpha = T\vec{J}\nabla\frac{\pi}{T} = T\vec{J}\left[\frac{1}{T}\nabla\pi - \frac{1}{T^2}\pi\nabla T\right] = \vec{J}[\nabla\pi - \alpha\nabla T] \tag{3}$$

For pure Thomson effect:

$$\vec{J}[\nabla\pi - \alpha\nabla T] = \vec{J}\left[\frac{d\pi}{dT} - \alpha\right]\nabla T = \tau\vec{J}\nabla T \tag{4}$$

With:

$$\tau = \frac{d\pi}{dT} - \alpha \tag{5}$$

The calculated efficiency is given by the figure of merit of a material as the following:

$$ZT = \frac{\sigma\alpha^2 T}{K} \tag{6}$$

The reduced efficiency coefficient is the following:

$$\eta_{r,d} = \frac{\sqrt{1+ZT} - 1}{\sqrt{1+ZT} + \frac{T_C}{T_h}}$$

The efficiency coefficient is the following:

$$\eta = \eta_{r,d}\frac{\Delta T}{T_h} \tag{7}$$

The relationship that links the two thermal ($K$) and electrical ($\sigma$) conductivities given by Weidman Franz [17] is used:

$$K = L\sigma T \tag{8}$$

Where $L = 2.14 \times 10^{-8}$ W.K$^{-2}$ is Lorenz number.

To complete our mathematical model, the initial and boundary conditions are used:

Initial conditions: $\forall t < t_0$

$T = T_0 = 305.15$ K

$V = V_0 = 0.14$ V

Boundary conditions: $\forall t > t_0$ (10)

At $x = 0, \forall y$: $T = T_C = 305.15$ K.

At $x = x_{max}, \forall y$: $T = T_h = 355.15$ K.

At $y = 0, \forall x$: the polyaniline sheet is considered adiabatic $Q = \left(\frac{dT}{dx}\right)_{y=0} = 0$.

At $y = y_{max}, \forall x$: the polyaniline sheet is considered adiabatic $Q = \left(\frac{dT}{dx}\right)_{y=y_{max}} = 0$.

High thermoelectric performance is related to maximizing the *ZT* merit factor. Achieving high performance of thermoelectric materials or high values of *ZT* requires materials having a high Seebeck coefficient as well as a high electrical conductivity, while maintaining a low value of thermal conductivity. This is quite difficult to achieve in the case of the conventional materials.

**3. Numerical techniques**

Our mathematical model coupled to the initial and boundary conditions are discretized using two numerical methods in order to be correctly analyzed. These two methods are the explicit finite difference method and the finite element method.

*3.1 Finite difference scheme*

The heat equation in the polyaniline sheet is schemed using the explicit form of the finite difference method. The resulting scheming equation is represented by the following form:

$$T_{(i,j)}^{n+1} = \left[MK_{(i+1,j)}^n - \frac{PT_{(i,j)}^n}{\rho\,Cp\,\Delta x}\right]T_{(i+1,j)}^n + \left[1 - K_{(i,j)}^n(2M+2N) + \frac{\Delta t}{\rho\,Cp}PT_{(i,j)}^n\left(\frac{1}{\Delta x}+\frac{1}{\Delta y}\right)\right]T_{(i,j)}^n +$$

$$MK_{(i-1,j)}^n T_{(i-1,j)}^n + \left[NK_{(i,j+1)}^n - \frac{PT_{(i,j)}^n}{\rho\,Cp\,\Delta y}\right]T_{(i,j+1)}^n + NK_{(i,j-1)}^n T_{(i,j-1)}^n + \frac{\Delta t}{\rho\,Cp}\frac{J_{(i,j)}^{2n}}{\sigma_{(i,j)}^n} \quad (11)$$

With $= \frac{\Delta t}{\rho Cp\Delta x^2}$ ; $N = \frac{\Delta t}{\rho Cp\Delta y^2}$ and $PT_{(i,j)}^n = \tau_{(i,j)}^n J_{(i,j)}^n$

The preliminary stability study of the program leads to a scheme of 100×100×60001 with a time step *Δt*=0.0006 s.

*3.2 Finite Element scheme*

The mathematical model is also solved using the finite element method. This last used the method of weighed residuals in the Galerkin formulation as reported by Dgheim [18,19]. Thus, the formulation of the weak integral of the thermoelectric problem is obtained as the following:

$$w(T,T^*) = \int_V T^* \rho C_p \dot{T} dV - \int_V T^* \nabla.(k\nabla T) dV - \int_V T^*\left(\frac{J^2}{\sigma_T} - \tau J\nabla T\right)dV = 0 \quad (12)$$

The finite element approximation of the equation (12) can be written as:

$$w(T,T^*) = \{T^*\}^T\left([D]\{\dot{T}\} + ([C]-[H]-[M])\{T\} + [L]\right) = 0 \quad (13)$$

Where, the dot represents differentiation with respect to time. The elementary matrices and vectors are given by:

$$[D] = \int_V [N]^T [\rho C_p][N] dV$$

$$[C] = \int_V [B]^T [K][B] dV$$

$$[H] = \int_{S\varphi} [N]^T [K][N] dS$$

$$[M] = \int_V [N]^T [\tau J][B] dV$$

$$[L] = \int_V [N]^T \left[\frac{J^2}{\sigma_T}\right] dV$$

[D] is the storage matrix, [C] is the conductive matrix on a volume, [H] is the conductive matrix in a surface, [M] is the Thomson vector and [L] is the constant matrix.

By using the finite difference scheme, equation (13) is written as the following:

$$[D]\frac{\{T\}^{n+1} - \{T\}^n}{\Delta t} + ([C] - [H] - [M])\{T\}^n + [L] = 0 \quad (14)$$

Then, the equation (14) is solved using an explicit finite difference method and Gauss-Legendre integration.

## 4. Results and Discussions

The importance of the ionic liquid polyaniline which is a conductive polymer is due to its exceptional electrical and electrochemical properties. Furthermore, its environmental stability, its easy polymerization and its low cost aroused a lot of interests.

*4.1 Model accuracy*

In order to validate our mathematical model, our results obtained by the finite difference method are compared to those obtained by the finite element one. Qualitative and quantitative agreements were observed between the two results (Figure 2).

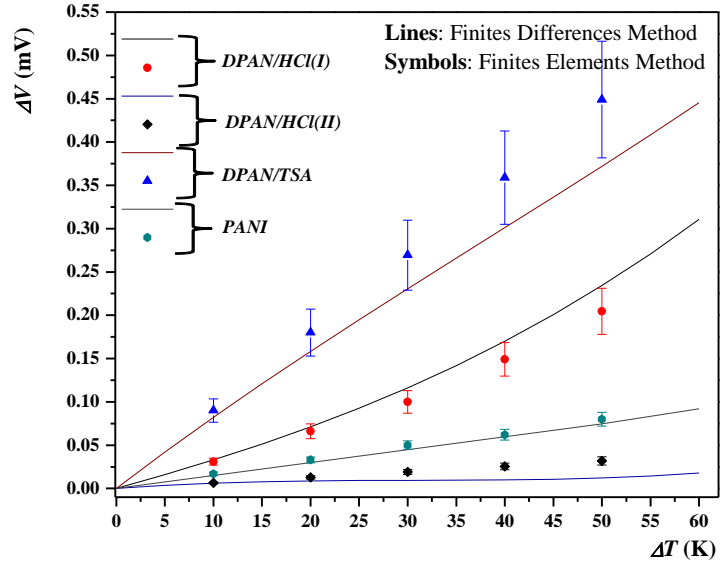

**Figure 2**: Comparison between the results of our model using the finite element method and the finite difference method for *PANI-DPAN/HCl(I)-DPAN/HCl(II)-DPAN/TSA*.

This agreement is observed for the non-doped PANI as well as for the various doped PANIs. The obtained relative error is less than 10%. Figure 2 represents the overall potential difference as a function of the temperature difference of the non-doped PANI and the three types of doped PANIs. The HCl(I) is a mineral dopant which is the hydrochloric acid of concentration 1.5 mol/l. The HCl(II) is a mineral dopant having higher concentration (> 1.5 mol/l). Finally, the toluenesulfonic acid (TSA) is an organic dopant. The potential difference obtained is 0.09 mV for the non-doped PANI, 0.45 mV for the DPAn/TSA, 0.20 mV for the DPAn/HCl(I) and 0.04 mV for the DPAn/HCl(II). Doping with a high concentration of HCl(II) showed a decrease in the voltage difference because its concentration exceeded a critical concentration. The two numerical methods are used to determine the potential difference values calculated for each value of the temperature difference $\Delta T$. Thus, one can deduce that the potential difference increases with the increase of the applied temperature difference.

*4.2 Our numerical results*

Polyaniline, like other conductive polymers, has strong basic centers. This polymer can be doped through an acid-base process, protonation. Then, the polyaniline can be protonated with an acid to give the corresponding polyaniline salt. This protonation makes the polymer more conductive. The doping of polyaniline leads to a very important modification of its electrical and thermal conductivities. The properties of the non-doped PANI and the HCl(I)-doped PANI are shown in Tables 1 and 2.

| Density $\rho$ (kg/m³) | Specific heat $Cp$ (J/(kg*K)) | Electrical conductivity $\sigma$ (S/m) | Thermal conductivity $K$ (W/(m.K)) | Seebeck Coefficient $\alpha$ (µv/K) |
|---|---|---|---|---|
| 1021 | 2140 | $10^{-2}$ | 0.17 | 2.98 |

**Table 1:** Properties of undoped polyaniline

| Density $\rho$ (kg/m³) | Specific heat $Cp$ (J/(kg*K)) | Electrical conductivity $\sigma$ (S/m) | Thermal conductivity $K$ (W/(m.K)) | Seebeck Coefficient $\alpha$ (µv/K) |
|---|---|---|---|---|
| 1245 | 2140 | 88 | 0.542 | 6.1732 |

**Table 2:** Properties of doped polyaniline *DPAn/HCl(I)*

The evolution of the temperature as a function of the length $x=4$ mm for different times is shown in Figure 3. One can notice that, the temperature increases along the sample and in terms of time to reach its maximum value equal to 355.15 K.

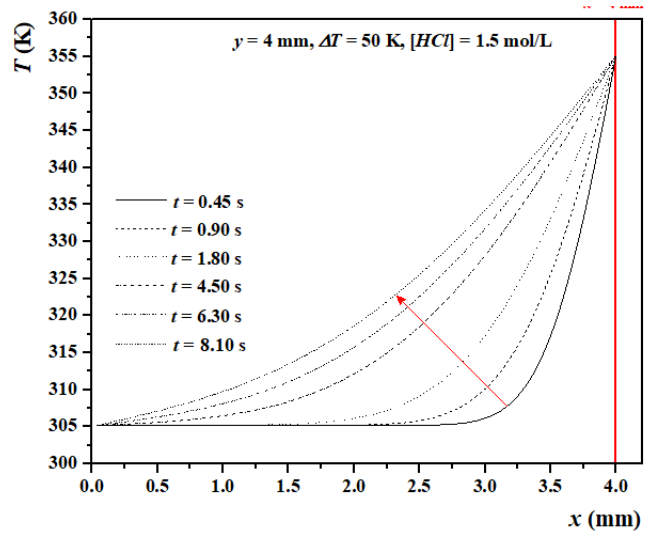

**Figure 3:** Variation of the temperature along the length of the cell for different times.

On the other hand, figure 4 represents the absolute value of the voltage variation in terms of the *x*-length of the HCl(I)-doped polyaniline, for different times. One can observe that the voltage difference increases to reach a maximum value of $2.3 \times 10^{-4}$ V at the position $x=4$ mm. The applied temperature difference generates a migration of positive and negative ions from one side to the other, resulting in the appearance of a voltage between the two faces of the used material. The doping effect accentuates strongly the delocalization of the electronic charges and, subsequently, the polymer becomes conductive.

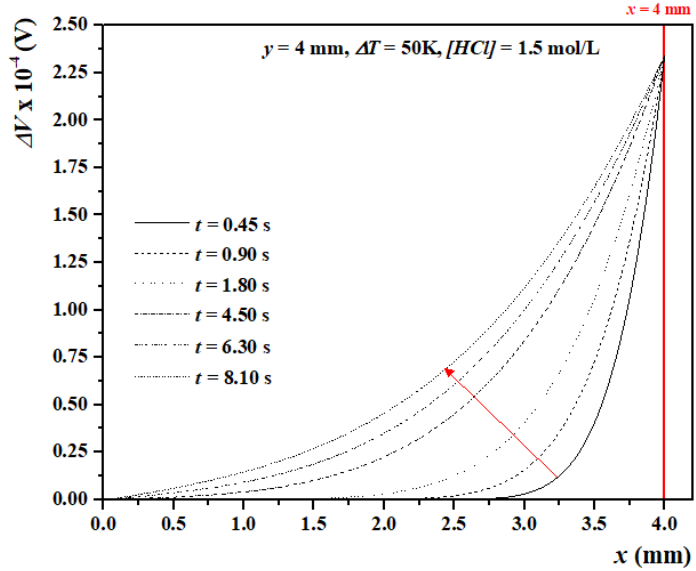

**Figure 4:** Variation of the voltage along the length of the cell for different times.

The Seebeck coefficient, which is one of the most interesting thermoelectric properties, is represented in Figure 5 as a function of the position $x$ in the sample.

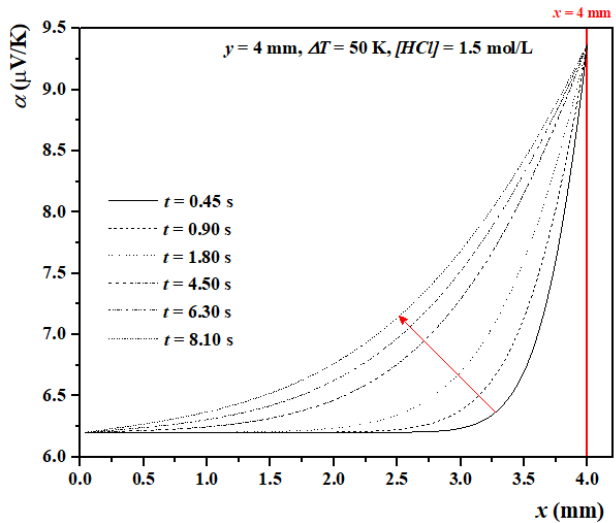

**Figure 5:** Variation of the Seebeck coefficient along the length of the cell for different times.

This coefficient evolves from an initial value of 6.25 µV/K for $x=0$ mm and increases along the length of the sample. This increase continues and reaches its maximum value of 9.25 µV/K for $x=4$ mm. The thermal gradient within the material generates differently positive and negative ions in order to create a potential difference. The Seebeck effect corresponds to the appearance of a voltage between the two faces of a material due to the non-uniform distribution of the ions in the ionic liquid, hence the relation: $\alpha = \Delta V / \Delta T$.

However, in order to study the performance of the thermoelectric cell, the evolution of the merit factor *ZT* along this cell is presented in figure 6. This factor *ZT* is proportional to the square of the Seebeck coefficient. The merit factor corresponding to the doped polyaniline, grows along the sample length and in terms of time. This factor takes a maximum value equal to $5.0 \times 10^{-6}$ for *x*=4 mm.

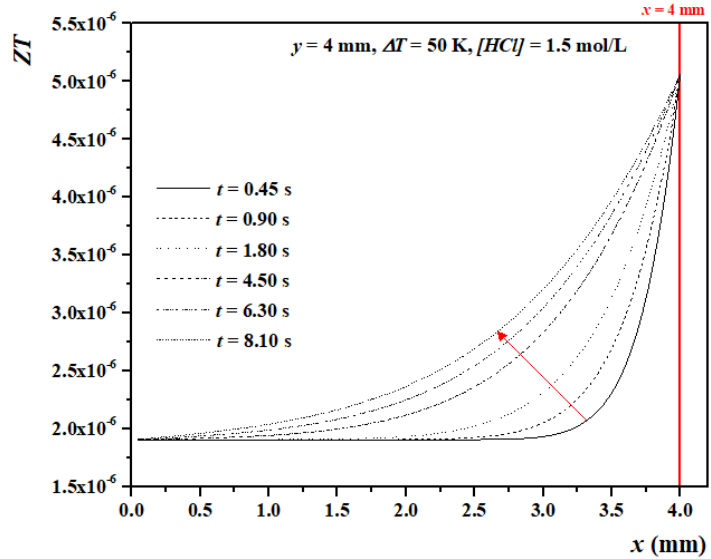

**Figure 6:** Variation of the merit factor along the length of the cell for different times.

Figure 7 shows the evolution of the Seebeck coefficient as a function of temperature for a doped PANI for different applied temperature differences. This coefficient increases gradually with increasing temperature differences. It varies between $\alpha = 6.5$ µV/K and $\alpha = 9.4$ µV/K. The increase in this coefficient shows an improvement in the sample efficiency. The high values of the Seebeck coefficient lead the charge carriers to move towards the cold side by canceling the induced Seebeck voltage. Thus, low carrier concentrations correspond to a high Seebeck coefficient. Electrons tend to diffuse better when a liquid is heated. If one end of the bar is hot, the electrons move faster and produce a flow to the colder side, negatively charged extremity. Therefore, this phenomenon improves the value of the potential difference.

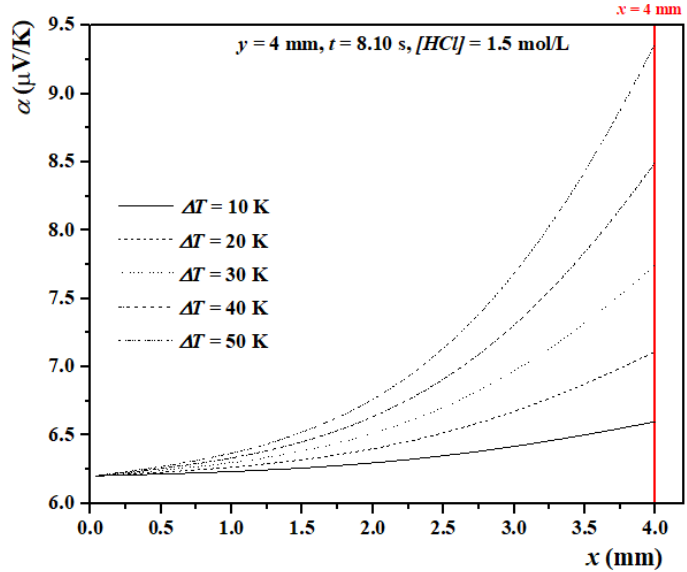

**Figure 7:** Variation of the Seebeck coefficient along the length of the cell for different values $\Delta T$.

The variation of the merit factor along the sample-length of a polyaniline doped with an HCl(I) for several temperature differences is shown in Figure 8. One can notice that the merit factor increases with the increase of the sample-length and the temperature differences $\Delta T$.

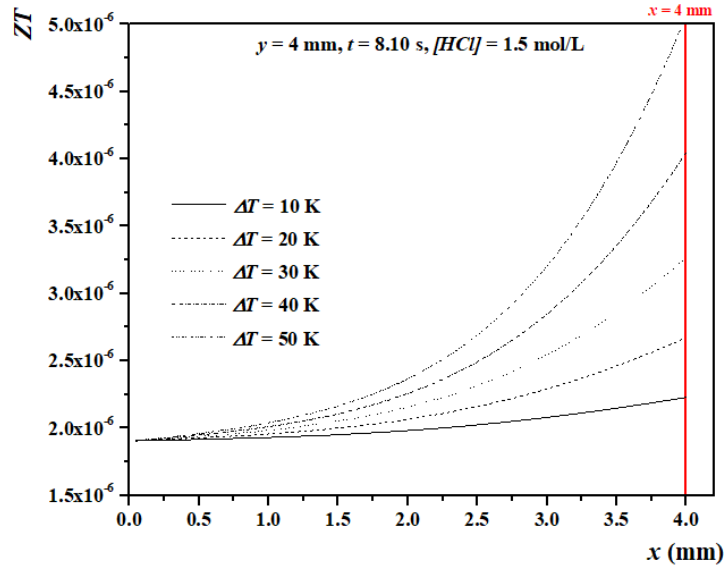

**Figure 8:** Variation of the merit factor along the length of the cell for different values $\Delta T$.

Figure 9 shows the changes in the Seebeck coefficient and the merit factor in terms of temperature. These two parameters gradually increase with the increasing of the temperature. The first parameter varies from $\alpha=6.2$ V/K to $\alpha=9.4$ V/K, and the second one reaches $ZT=5.06\times10^{-6}$. This increasing shows an improvement of the sample efficiency.

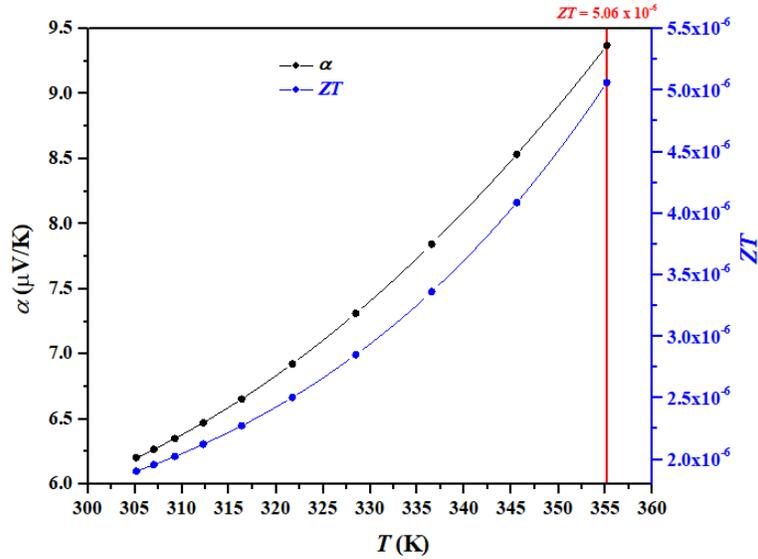

**Figure 9:** Variation of the Seebeck coefficient and the merit factor as functions of the temperature.

The efficiency and reduced efficiency depending on the merit factor are presented in Figure 10. These efficiencies increase with the increase of the temperature and reach $1.9\times10^{-7}$ and $1.38\times10^{-6}$ respectively at a temperature $T=355.15$ K.

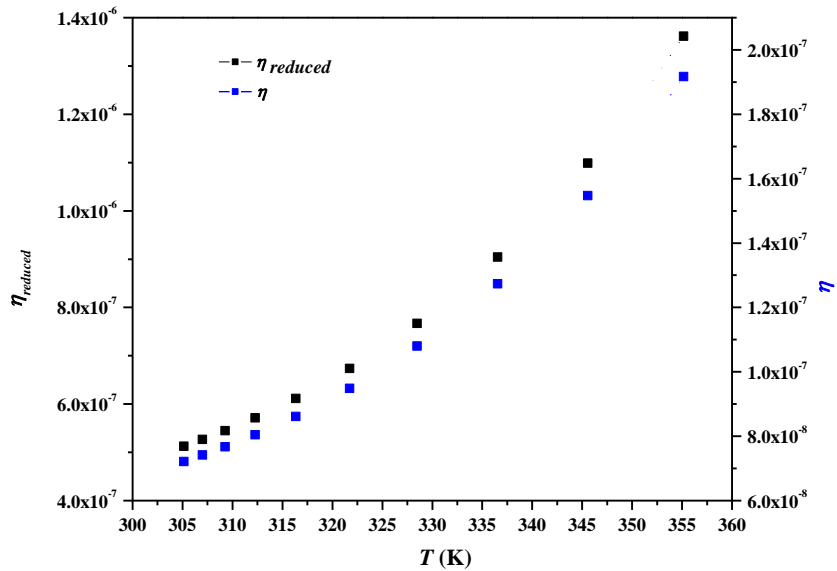

**Figure 10:** Evolution of the efficiency and reduced efficiency as a function of the temperature.

Figure 11 depicts the Seebeck coefficient as a function of the temperature for the three doped PANIs. One can notice that the Seebeck coefficient of DPAn/HCl(I) increases from 6 $\mu V/K$ ($T=305$ K) to 30 $\mu V/K$ ($T=480$ K). Concerning the DPAn/HCl(II), it increases from 2 $\mu V/K$ ($T=305$ K) to 10 $\mu V/K$ ($T=480$ K). In the same way, the DPAn/TSA decreases from 17.5 $\mu V/K$ to 15 $\mu V/K$ ($T=350$ K), then increases again to 25 $\mu V/K$ ($T=480$ K).

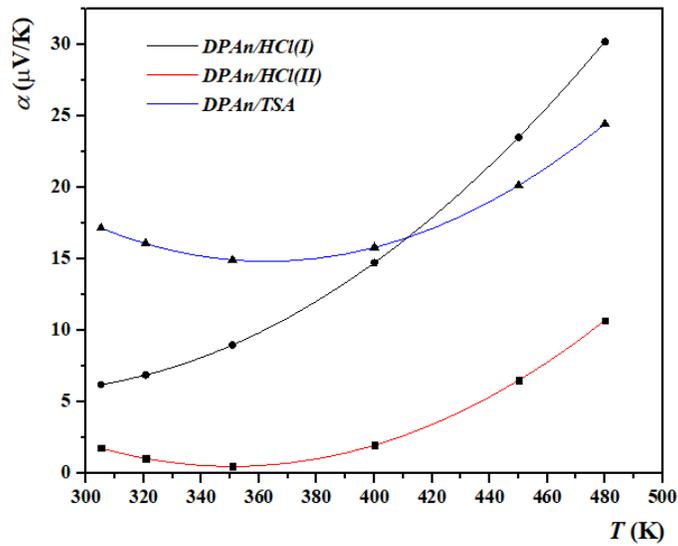

**Figure 11:** Variation of the Seebeck coefficient as a function of temperature for the three different dopings.

Figure 12 represents the merit factor $ZT$ as a function of temperature for the three doped PANIs. The merit factor of the DPAn/HCl(I) increases from 0 ($T=305$ K) to $6.5\times10^{-5}$ ($T=480$ K). For the DPAn/HCl(II), it increases from 0 ($T=305$ K) to $1.0\times10^{-4}$ ($T=480$ K). For the DPAn/TSA, it decreases from $7.5\times10^{-5}$ ($T=305$ K) to $5.0\times10^{-5}$ ($T=350$ K), then increases to $2.3\times10^{-4}$ ($T=480$ K). This improvement in the merit factor is due to the decrease in the thermal conductivity and the increase in the electrical conductivity.

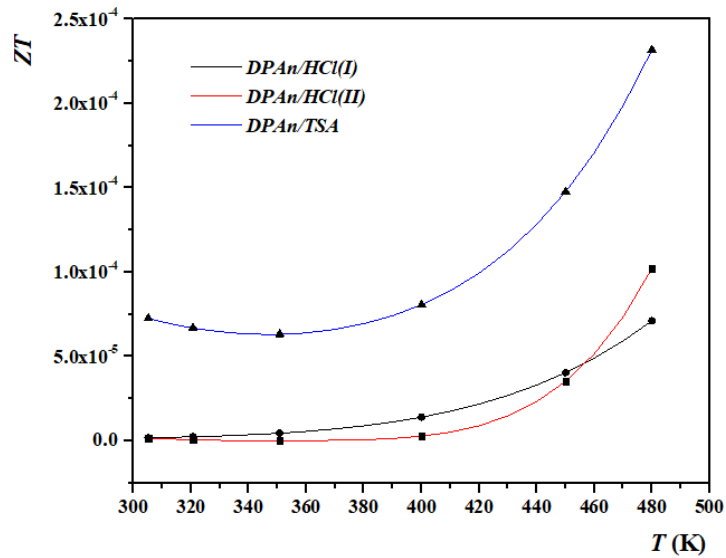

**Figure 12:** Variation of the merit factor as a function of temperature for the three different dopings.

Finally, one can demonstrate that for different dopants, the doped polyaniline which has a very high dopant concentration DPAn/HCl(II) is less efficient than the DPAn/HCl(I) since the polymer was deteriorated. While DPAn/TSA organic dopant performed better than DPAn/HCl inorganic dopants.

## 5   Conclusions

A numerical model describing the heat conversion into electricity that developed in the thermoelectric ionic liquid such as polyaniline is determined. Our mathematical model introduces the temperature difference applied at the ends of the sample to treat the electric proprieties of the polyaniline sheet. The results of our both numerical techniques are compared and good qualitative and quantitative agreements are observed. The obtained results are as follows:

 - The doping of the PANI ionic liquid presents good results and the DPAn/HCl(I) is more efficient than the PANI without doping.
 - The PANI having a very high dopant concentration DPAn/HCl(II) is less efficient than the DPAn/HCl(I), since quantity of doping is very high and the polymer is deteriorated. While DPAn/TSA organic acid dopant performed better than DPAn/HCl inorganic dopant.
 - For DPAn/TSA, the potential difference found is 0.45 mV at a temperature difference of 50 K while for the DPAn/HCl(I), the value of the potential difference obtained is equal to 0.20 mV.

**Acknowledgment**

This work is supported by a grant from the Lebanese University.